\documentclass[pdflatex,sn-mathphys-num]{sn-jnl}
\usepackage[T1]{fontenc}
\usepackage[utf8]{inputenc}
\usepackage{graphicx}
\graphicspath{{results/}}
\usepackage[version=4]{mhchem}
\usepackage{booktabs,tabularx,array}
\usepackage{multirow}
\usepackage{makecell}        

\usepackage{amsmath,amssymb,bm}
\usepackage{xcolor}
\usepackage{siunitx}
\usepackage{placeins}

\usepackage{color} 

\definecolor{sqdrevisionred}{RGB}{190,0,0}

\newcommand{\sqdEh}{E_{\mathrm h}}

\begin{document}

\title[Ground-state energies via SQD]{Ground-State Energy Estimation of HeH$^{+}$, ArH$^{+}$, and H$_2$O via Sample-Based Quantum Diagonalization}

\author[1]{\fnm{Jubin} \sur{Park}}
\author[1]{\fnm{Chae-Hyun} \sur{Yoon}}
\author[1]{\fnm{Minkyu} \sur{Lee}}
\author*[1]{\fnm{Myung-Ki} \sur{Cheoun}}\email{cheoun@ssu.ac.kr}

\affil[1]{\orgdiv{Department of Physics and Origin of Matter and Evolution of Galaxies (OMEG) Institute}, \orgname{Soongsil University}, \orgaddress{\city{Seoul}, \postcode{06978}, \country{Republic of Korea}}}

\abstract{
Accurate ground-state energies are essential for understanding molecular structure, chemical bonding, and reaction energetics in quantum chemistry. In this work, we investigate the ground-state properties of the molecular systems HeH$^+$, ArH$^+$, and H$_2$O using Sample-Based Quantum Diagonalization (SQD), a hybrid quantum-classical framework designed for near-term quantum devices. Unlike variational approaches such as the Variational Quantum Eigensolver (VQE), which require deep parameterized circuits and repeated expectation-value measurements, SQD reconstructs a
low-energy determinant subspace directly from measured bitstrings.

For the present calculations, bitstrings were generated on IBM
quantum hardware using shallow local unitary cluster Jastrow
(LUCJ) circuits whose parameters were constructed from the
$t_1$ and $t_2$ amplitudes of coupled-cluster singles and doubles (CCSD) calculations based on restricted Hartree--Fock (RHF) references.
From these samples, we compute ground-state potential-energy curves of HeH$^+$, ArH$^+$, and H$_2$O with the 6-31G and cc-pVDZ basis sets. For all three systems, the SQD results obtained with the cc-pVDZ basis closely follow the corresponding same-basis CCSD energies and reproduce the equilibrium-region trends of the potential-energy curves. 
HeH$^+$ and ArH$^+$ were chosen as simple yet astrophysically important molecular-ion benchmarks, while H$_2$O was included as a representative polyatomic molecule to assess the applicability of SQD beyond diatomic ionic systems. {At the adopted equilibrium geometries, the deviations from the same-active-space CASCI (complete active space configuration interaction) references are 0.00, 2.51, and 6.34 mHa
for HeH$^+$, ArH$^+$, and H$_2$O, respectively.}

These results demonstrate the feasibility of hardware-assisted SQD
for the present benchmark systems and motivate further studies of
its accuracy and computational scaling for larger molecular active
spaces.
}


\maketitle

\section{Introduction}\label{section1}
Understanding the ground-state properties of small molecular systems is of fundamental importance in quantum chemistry and astrophysics.
Among such systems, the helium hydride ion (HeH$^{+}$) and the argon hydride ion (ArH$^{+}$) occupy a unique position due to their distinct roles in cosmic chemical evolution.
HeH$^{+}$ is widely regarded as the first molecular species formed in the early universe, emerging shortly after recombination via radiative association processes~\cite{HeH_astro,HeH_arxiv}.
Its formation, stability, and destruction pathways play a crucial role in primordial chemistry and significantly influence the thermal and chemical evolution of the early cosmos~\cite{HeH_arxiv,HeH_Palla}.
In contrast, ArH$^{+}$, although not stable under typical terrestrial conditions, has been firmly detected in diffuse interstellar media, where it serves as a sensitive tracer of ionized and low-density astrophysical environments~\cite{ArH_Schilke,ArH_Muller}.
Accurate theoretical descriptions of these ionic molecules are therefore essential for interpreting astronomical observations and for validating models of molecular formation under extreme conditions.

Obtaining reliable ground-state energies for such ionic systems can still be computationally demanding.
High-level classical electronic-structure methods, including coupled-cluster and explicitly correlated approaches, are capable of delivering benchmark-quality results, but their computational cost grows rapidly with increasing basis-set size and correlation complexity~\cite{accuracy_scaling,explicitly_correlated}.
This scaling severely limits their applicability to larger or more chemically realistic systems.
In recent years, quantum computing has emerged as a promising alternative paradigm for electronic-structure calculations, offering, in principle, a more favorable scaling for many-body problems~\cite{NISQ,VQE_Peruzzo}.
However, many quantum algorithms currently under active development-most notably variational approaches-face significant obstacles on noisy intermediate-scale quantum (NISQ) devices, primarily due to the need for deep ansatz circuits and extensive measurement overhead~\cite{VQE_overhead}.

Sample-Based Quantum Diagonalization (SQD) provides a conceptually distinct and practically advantageous route to addressing these challenges, particularly when contrasted with variational quantum algorithms such as the variational quantum eigensolver (VQE)~\cite{VQE_overhead,SQD_Barison}.
Whereas VQE relies on repeated evaluations of expectation values within deep, parameterized ansatz circuits-an approach that is highly susceptible to noise accumulation and measurement overhead on noisy intermediate-scale quantum (NISQ) devices-SQD ~\cite{VQE_practice} avoids the repeated on-hardware energy minimization required in conventional VQE workflows.
Instead, SQD reconstructs a low-energy subspace directly from samples obtained using relatively shallow quantum circuits, followed by classical diagonalization of the Hamiltonian within this reduced space~\cite{SQD_embedding,SC_diag,SQD_MA}.

This sample-based subspace construction naturally aligns with the capabilities and limitations of current quantum hardware, offering enhanced robustness to noise and a substantial reduction in measurement costs~\cite{Noise_QAlgorithm, VQE_moleculenoise}.
By operating on bitstring samples rather than expectation values, SQD mitigates the sensitivity to gate errors and statistical fluctuations that often limit the performance of variational methods~\cite{VQE_Verteletskyi}.
Moreover, by efficiently identifying and exploiting the most physically relevant regions of Hilbert space, SQD enables chemically accurate ground-state energy calculations without requiring exhaustive sampling of the full configuration space~\cite{SQD_MA}, thereby providing a scalable and reliable alternative to variational approaches for quantum chemistry applications on near-term devices~\cite{VQE_SQD}.

In this work, we apply the SQD framework to the molecular systems HeH$^+$, ArH$^+$, and H$_2$O. HeH$^+$ and ArH$^+$ are of clear astrophysical interest ~\cite{HeH_astro,ArH_detec}, while H$_2$O provides a representative polyatomic benchmark that allows us to assess the applicability of SQD beyond diatomic ionic molecules. By considering these systems together, we demonstrate both the physical relevance and the methodological flexibility of SQD-based quantum simulations. More broadly, our results support further investigation of hardware-assisted SQD as a route from simple molecular ions to larger polyatomic active spaces. The distinctive contribution of the present work is a hardware-executed SQD study of one-dimensional potential-energy scans for two astrophysically relevant molecular ions and a
46-qubit polyatomic benchmark, combined with same-active-space CASCI and CCSD comparisons and explicit reporting of circuit and sampling resources.

The remainder of this paper is organized as follows. Section 2 describes the theoretical framework of the Sample-Based Quantum Diagonalization method and outlines the computational details used in this study. Section 3 presents the calculated ground-state energies and potential energy curves of HeH$^{+}$, ArH$^{+}$, and H$_2$O, together with a discussion of the numerical results. Finally, Section 4 summarizes the main conclusions and discusses the implications of the present work. 

\section{Method}\label{sec:2}
\subsection{Electronic Structure Framework}

In this section, we briefly summarize the Sample-Based Quantum Diagonalization (SQD) framework introduced in Refs.~~\cite{Kry_SQD,SC_diag,SQD_embedding}, which forms the basis of the present calculations.
Within the Born-Oppenheimer approximation, the electronic ground state of a molecular system is obtained by solving the time-independent Schr\"odinger equation, $\hat{H_e}\lvert{\Psi_0}\rangle = E_0\lvert{\Psi_0}\rangle$, where $\hat{H_e}$ is the electronic Hamiltonian and $E_0$ denotes the ground-state energy.
In a finite orthonormal spatial-orbital basis
$\{\phi_p\}$, the clamped-nuclei molecular Hamiltonian is expressed
in second quantization as
\begin{equation}
\hat{H_e}
=
E_{nuc}
+
\sum_{pq\sigma}
h_{pq}\,
\hat{a}^{\dagger}_{p\sigma}
\hat{a}_{q\sigma}
+
\frac{1}{2}
\sum_{pqrs}
\sum_{\sigma\tau}
h_{pqrs}
\hat{a}^{\dagger}_{p\sigma}
\hat{a}^{\dagger}_{q\tau}
\hat{a}_{s\tau}
\hat{a}_{r\sigma},
\label{eq:1}
\end{equation}
{\begin{equation}
h_{pq} =
\int
\phi_p^{*}(\mathbf{r})
\left(
-\frac{1}{2}\nabla^2
-
\sum_{\alpha}
\frac{Z_{\alpha}}{|\mathbf{r}-\mathbf{R}_{\alpha}|}
\right)
\phi_q(\mathbf{r})
\, d\mathbf{r}
\end{equation}
\begin{equation}
h_{pqrs} =
\iint
\phi_p^{*}(\mathbf{r}_1)
\phi_q^{*}(\mathbf{r}_2)
\frac{1}{|\mathbf{r}_1-\mathbf{r}_2|}
\phi_r(\mathbf{r}_1)
\phi_s(\mathbf{r}_2)
\, d\mathbf{r}_1 d\mathbf{r}_2
\end{equation}
}
where $p,q,r,s$ label spatial orbitals, $E_{nuc}$ is the nuclear repulsion energy, {$h_{pq}$} and $h_{pqrs}$ are one- and two-electron integrals, respectively, and the fermionic operators satisfy canonical anticommutation relations.
Here, $\hat{a}_{p\sigma}^{\dagger}$ and $\hat{a}_{q\sigma}$ denote the fermionic creation and annihilation operators that create and remove an electron with spin $\sigma$ in the spin orbital $p\sigma$ and $q\sigma$, respectively.

The fermionic Hamiltonian is mapped to qubits using the Jordan-Wigner transformation~~\cite{JW}, $\hat{H_q}=\sum_i \alpha_i P_i$, where $P_i$ are Pauli strings acting on qubits and $\alpha_i \in \mathbb{R}$.
In this representation, computational basis states $\lvert{x}\rangle \in \{0,1\}^{N_q}$ correspond directly to occupation-number Slater determinants.

The exact ground state admits the expansion $\lvert \Psi_0\rangle = \sum_x c_x\lvert x \rangle$, where the dimension of the Hilbert space grows combinatorially with the number of spin orbitals.
Consequently, exact diagonalization rapidly becomes intractable for realistic active spaces.

\subsection{LUCJ Ansatz}
A truncated Local Unitary Cluster Jastrow (LUCJ)~\cite{LUCJ,LUCJ_orbital_rotation} ansatz was chosen because it provides a compact representation of electron correlation while maintaining shallow circuit depth, which is particularly advantageous for near-term quantum hardware.
Compared with more conventional ansatzes such as unitary coupled-cluster or hardware-efficient circuits, the LUCJ ansatz offers a favorable balance between expressibility and circuit complexity.
This property is especially beneficial within the SQD framework, where shallow circuits enable efficient sampling of physically relevant determinants while reducing the impact of noise and measurement overhead on NISQ devices.
We employ a truncated LUCJ ansatz acting on the restricted Hartree-Fock reference state:
\begin{equation}
\lvert \Psi(\theta) \rangle
=
\prod_{\mu=1}^{L}
e^{\hat{\kappa}_\mu}
\, e^{i \hat{J}_\mu (\theta)}
\, e^{-\hat{\kappa}_\mu}
\lvert \Psi_{\mathrm{RHF}} \rangle .
\label{eq:2}
\end{equation}
Here, $\hat{\kappa}_\mu$ generates orbital rotations, while
$\hat{J}_\mu$ is the spin-balanced diagonal-Coulomb operator
\begin{equation}
\hat{J}_\mu
=
\frac{1}{2}
\sum_{p,q=0}^{N_{\rm orb}-1}
\sum_{\sigma,\tau\in\{\alpha,\beta\}}
J_{pq}^{(\mu;\sigma\tau)}
\hat n_{p\sigma}
\hat n_{q\tau},
\qquad
\hat n_{p\sigma}
=
\hat a_{p\sigma}^{\dagger}
\hat a_{p\sigma}.
\end{equation}
The factor $1/2$ avoids double counting under the unrestricted
sum over $p$ and $q$. For the spin-balanced parametrization,
$J^{(\mu;\alpha\alpha)}=J^{(\mu;\beta\beta)}$ and
$J^{(\mu;\alpha\beta)}=J^{(\mu;\beta\alpha)}$, with symmetric
orbital-index matrices. The quantity $L$ denotes the number of
LUCJ ansatz repetitions and is distinct from the
post-transpilation circuit depth.

Unlike variational quantum eigensolvers, the parameters are not iteratively optimized to minimize energy on hardware.
Instead, the circuit serves as a structured sampler designed to amplify physically relevant determinants while maintaining shallow depth and reduced susceptibility to noise accumulation.

Measurement in the computational basis produces bitstrings that are later used for subspace construction within the SQD procedure. 

In the present calculations, the spin-balanced LUCJ parameters
were initialized independently at each molecular geometry from
the RHF-based CCSD $t_1$ and $t_2$ amplitudes.
The amplitudes were mapped to the LUCJ form by double
factorization using $n_{\rm reps}=1$ and
\texttt{optimize=False}.
Thus, only one LUCJ repetition was retained, and no
subsequent numerical optimization of the truncated factorization
was performed.

The diagonal-Coulomb interactions were restricted to same-spin
nearest-neighbor pairs
$\{(p,p+1)\}_{p=0}^{N_{\rm orb}-2}$
and opposite-spin on-site pairs
$\{(p,p)\}_{p=0}^{N_{\rm orb}-1}$.
Consequently, the prepared state is a truncated LUCJ
approximation initialized from CCSD amplitudes, rather than an
exact representation of the CCSD wavefunction.
No iterative variational optimization was performed on the
quantum hardware.

\subsection{Sample-Based Quantum Diagonalization for Ground-State Energy}
Sample-Based Quantum Diagonalization (SQD) approximates the solution of the Schr\"odinger equation by constructing a reduced Hilbert subspace that captures the dominant contributions to the low-energy spectrum.
Instead of evaluating expectation values of the Hamiltonian, SQD reconstructs a physically relevant determinant manifold directly from samples obtained from a shallow quantum circuit.

After preparing the trial state $\lvert \Psi(\theta) \rangle$ using the LUCJ ansatz, repeated measurements in the computational basis generate a collection of bitstrings distributed according to the probability $p(x)=| \langle x \lvert \Psi(\theta) \rangle |^2$.
From $N_s$ measurements, we obtain a multiset of sampled configurations~~\cite{SC_diag}
\begin{equation}
\tilde{X} = \{ x^{(1)}, x^{(2)}, \dots, x^{(N_s)} \}.
\label{eq:3}
\end{equation}
{
{Since multiple measurements may yield identical bitstrings,
identical outcomes are aggregated into unique configurations
together with their empirical probabilities. After the
configuration-recovery and postselection steps described in
Sec.~2.4, the resulting physically admissible configuration set is
denoted
$
\mathcal X=\{x_1,x_2,\ldots,x_D\}.
$
}
These determinants define a reduced Hilbert subspace that is expected to contain the dominant contributions to the ground-state wavefunction.
The projection operator onto this sampled subspace is
\begin{equation}
P_X = \sum_{i=1}^D \lvert x_i \rangle \langle x_i \lvert.
\label{eq:5}
\end{equation}
Using this projector, the qubit Hamiltonian $\hat{H}_q$ is restricted to the sampled determinant manifold,
\begin{equation}
\tilde{H} = P_X \hat{H_q} P_X.
\label{eq:6}
\end{equation}
The approximate ground-state energy is then obtained by solving the projected eigenvalue problem
\begin{equation}
\tilde{H}\bf{c} = \tilde{E}_0 \bf{c},
\label{eq:7}
\end{equation}
with the corresponding approximate wavefunction
\begin{equation}
\lvert \tilde{\Psi}_0 \rangle = \sum_{i=1}^D c_i \lvert x_i \rangle.
\label{eq:8}
\end{equation}
This procedure replaces the stochastic estimation of Hamiltonian expectation values with deterministic classical diagonalization within a physically motivated configuration subspace.

To further improve numerical stability and mitigate statistical fluctuations arising from finite sampling, the measurement data are used to generate $B$ subsampled batches from a common empirical measurement distribution
\begin{equation}
\tilde{X}^{(b)}, b=1,...,B.
\label{eq:9}
\end{equation}
Each subsample generates its own projected Hamiltonian $\tilde{H}^{(b)}$ and corresponding ground-state estimate $\tilde{E}_0^{(b)}$.
The final SQD energy estimator is defined as {$\tilde{E}_0 = \displaystyle \min_{b} \tilde{E}_0^{(b)}$.}
This multi-batch strategy reduces the sensitivity of the final estimate to any single
subsample and improves robustness against statistical fluctuations in the measurement data.

\subsection{Electronic Configuration Recovery}
\label{sec:config_recovery}

Due to hardware noise and imperfect state preparation, the raw
sampled set $\widetilde{X}$ may contain bitstrings whose
spin-resolved Hamming weights do not match the target electron
numbers.
A simple postselection procedure would discard all such samples.
In the present calculations, we instead use the self-consistent
configuration-recovery procedure of Ref.~\cite{SC_diag}, which
attempts to recover part of their information before constructing
the determinant subspaces.

In the initial recovery step, before an updated SQD occupation
profile is available, the measured distribution is postselected
according to the target spin-resolved particle numbers.
In subsequent iterations, the current spin-orbital occupations,
$\{n_{p\sigma}^{(k)}\}$, are used to guide probabilistic bit flips
for configurations with incorrect Hamming weights.
The original and recovered configurations are then postselected
to satisfy

\begin{equation}
N_{\alpha}(x)
=
N_{\alpha}^{\rm target},
\qquad
N_{\beta}(x)
=
N_{\beta}^{\rm target},
\qquad
x\in X_R .
\label{eq:10}
\end{equation}

The resulting empirical distribution is subsampled into $B$
batches.
For each batch, the Hamiltonian is projected onto the corresponding
determinant subspace and diagonalized classically.
The spin-orbital occupations obtained from the lowest-energy batch
are used to update the recovery model for the next iteration,

\[
n_{p\sigma}^{(k)}
=
\left\langle
\widetilde{\Psi}_0^{(k)}
\right|
\hat n_{p\sigma}
\left|
\widetilde{\Psi}_0^{(k)}
\right\rangle .
\]

Convergence is declared when both the minimum batch energy and the
spin-resolved orbital occupations stabilize,

\begin{equation}
\left|
E^{(k+1)}
-
E^{(k)}
\right|
<
\varepsilon_E,
\qquad
\max_{p,\sigma}
\left|
n_{p\sigma}^{(k+1)}
-
n_{p\sigma}^{(k)}
\right|
<
\varepsilon_n .
\label{eq:11}
\end{equation}
{Here, $E^{(k)}=\min_b E_0^{(k,b)}$ denotes the minimum batch energy
at recovery iteration $k$. This iterative recovery--postselection--subsampling procedure
refines the measured determinant distribution toward a compact,
physically admissible subspace and improves the numerical stability
of the subsequent selected-configuration-interaction (SCI)
diagonalization.}

\subsection{Projected Hamiltonian and Subspace Diagonalization}
Within the recovered configuration space $X_R$, the Hamiltonian matrix elements are defined as
\begin{equation}
H_{ij}
=
\langle x_i \lvert \hat{H_q} \rvert x_j \rangle,
\qquad
x_i, x_j \in \mathcal{X}_R .
\label{eq:12}
\end{equation}
Because the electronic Hamiltonian contains at most two-body
fermionic operators, nonzero matrix elements arise only between
determinants related by at most a double excitation, i.e., by at
most two spin-orbital substitutions. In the occupation-bit
representation, this corresponds to a Hamming distance of at most
four. Consequently, the projected Hamiltonian matrix is sparse, allowing efficient classical diagonalization even when the configuration subspace becomes moderately large. 
Diagonalization of this projected Hamiltonian yields approximate eigenvalue ${E}_{\mu}$ and eigenvectors
\begin{equation}
\lvert \tilde{\Psi}_{\mu} \rangle
=
\sum_{x_i \in \mathcal{X}_R}
c_i^{(\mu)}
\lvert x_i \rangle .
\label{eq:13}
\end{equation}
The lowest eigenvalue provides an estimate of the ground-state energy, $\tilde{E}_0 = \min_\mu E_\mu$.
To improve robustness against finite-sampling fluctuations, the diagonalization is repeated for multiple subsampled configuration sets generated from the same empirical measurement pool, and the minimum batch energy is used as the final SQD estimator. This step yields the approximate ground-state energy within the sampled determinant subspace.

\subsection{Computational details}

The computational parameters employed in the present SQD simulations are summarized in Table~1. For the HeH$^+$ molecular ion, all orbitals generated by the chosen basis sets were included in the active space because the system contains only two electrons, and therefore no frozen-core approximation was applied. In contrast, for the ArH$^+$ molecule, five core orbitals associated with the inner-shell electrons of the Ar atom were frozen in order to reduce the size of the active space while preserving the chemically relevant valence correlation. For H$_2$O, one oxygen core orbital was frozen, and the remaining orbitals were included in the active space used for the SQD calculations. 
Under this configuration, four $\alpha$ and four $\beta$ electrons were explicitly treated in the electronic-structure calculations.
The resulting active spaces consist of 12 spatial orbitals (24 spin orbitals) for the 6-31G basis set and 23 spatial orbitals (46 spin orbitals) for the cc-pVDZ basis set.
The equilibrium geometry of H$_{2}$O with an O-H bond length of 0.95782 {\AA} and an H--O--H bond angle of  $104.485^\circ$ was used \cite{cc_pvdz,ccsd, ccpvdz_basis1}.

The quantum circuits used for the sampling procedure were executed on the IBM Quantum superconducting processor \texttt{ibm\_torino}, which is based on the Heron r1 architecture and provides up to 133 physical qubits. The shallow circuits associated with the truncated LUCJ ansatz were implemented on this device to generate measurement bitstrings distributed according to the corresponding measurement probabilities. These sampled configurations were subsequently used to construct the reduced determinant subspace employed in the SQD diagonalization procedure.

Table~1 summarizes the active-space definitions and qubit requirements for the three molecular systems considered in this work. The inclusion of H$_2$O extends the present study beyond diatomic ionic molecules to a representative polyatomic benchmark, thereby allowing us to assess the {applicability} of the SQD framework in a more general molecular setting.

Depending on the basis set employed, the number of spatial orbitals ranges from 4 to 23, corresponding to 8--46 spin orbitals after the fermion-to-qubit mapping. Within the Jordan--Wigner transformation, each spin orbital is mapped to one qubit, so that the number of qubits used in the SQD simulations is equal to the number of spin orbitals listed in Table~1. The active electron configuration corresponds to one $\alpha$ and one $\beta$ electron for HeH$^+$, four $\alpha$ and four $\beta$ electrons for ArH$^+$, and four $\alpha$ and four $\beta$ electrons for H$_2$O after the frozen-core approximation.

In our calculations, the number of qubits corresponds directly to the number of spin orbitals in the active space after the fermion-to-qubit mapping. Consequently, the largest simulations considered here involve up to 46 qubits for the H$_2$O system with the cc-pVDZ basis. In the SQD procedure, the Hamiltonian is diagonalized within a
recovered SCI subspace of the
fixed-$(N_\alpha,N_\beta)$ active-space sector. The dimension of the
corresponding full determinant space is

\begin{equation}
D_{\rm FCI}
=
\binom{N_{\rm orb}}{N_\alpha}
\binom{N_{\rm orb}}{N_\beta},
\end{equation}
where $N_{\rm orb}$ denotes the number of active spatial orbitals. The calculations were performed in the $M_S=0$ sector fixed by
$N_\alpha=N_\beta$. No explicit spin projection was imposed. Because the recovered SCI-subspace dimension depends on the
subsampling batch and configuration-recovery iteration, no single
intermediate subspace dimension is reported in Table~1.
The run-level settings listed in Table~1(b) refer specifically to
the H$_2$O/cc-pVDZ hardware scan.

{The CASCI reference energies were obtained by full
configuration-interaction diagonalization of the same frozen-core
active-space Hamiltonians used in the SQD calculations, within the
fixed-$(N_\alpha,N_\beta)$ sector. The calculations were performed
using the PySCF CASCI implementation with its built-in full configuration interaction (FCI) solver with an energy-convergence threshold of
{1 × 10$^{-8} E_h$} and a maximum of {200} solver iterations.
The RHF orbitals, one- and two-electron integrals, frozen-core
constants, and nuclear-repulsion energies were identical to those
used in the corresponding SQD calculations.}

\begingroup
\setlength{\tabcolsep}{3.0pt}
\setlength{\aboverulesep}{1.0pt}
\setlength{\belowrulesep}{1.0pt}
\renewcommand{\arraystretch}{0.92}

\begin{table*}[!t]  
\centering
\caption{Computational specifications and recorded quantum-execution
parameters used in the sample-based quantum diagonalization (SQD)
calculations. Panel (a) lists the active-space definitions for all
molecule/basis combinations. Panel (b) reports the hardware,
circuit-resource, sampling, and SQD post-processing parameters for
the H$_2$O/cc-pVDZ scan. Unless otherwise stated, the sampling and SQD post-processing parameters listed in Table~1(b) were also used for the remaining molecule/basis combinations. }
\label{tab:sqd_computational_hardware_parameters}

\vspace{1pt}
\fontsize{6.6}{7.2}\selectfont
\noindent\textbf{(a) Molecular active spaces}\par
\vspace{1pt}
\begin{tabularx}{\textwidth}{@{}l l >{\centering\arraybackslash}p{0.145\textwidth} >{\centering\arraybackslash}p{0.145\textwidth} >{\centering\arraybackslash}p{0.105\textwidth} >{\centering\arraybackslash}p{0.055\textwidth} >{\centering\arraybackslash}p{0.055\textwidth}@{}}
\toprule
Molecule & Basis & \makecell{Frozen-core\\spatial orbitals} & \makecell{Active spatial\\orbitals, $N_{\mathrm{orb}}$} & \makecell{Spin-orbital\\qubits, $N_q$} & $N_{\alpha}$ & $N_{\beta}$ \\
\midrule
HeH$^+$ & 6-31G   & 0 & 4  & 8  & 1 & 1 \\
HeH$^+$ & cc-pVDZ & 0 & 10 & 20 & 1 & 1 \\
ArH$^+$ & 6-31G   & 5 & 10 & 20 & 4 & 4 \\
ArH$^+$ & cc-pVDZ & 5 & 18 & 36 & 4 & 4 \\
H$_2$O  & 6-31G   & 1 & 12 & 24 & 4 & 4 \\
H$_2$O  & cc-pVDZ & 1 & 23 & 46 & 4 & 4 \\
\bottomrule
\end{tabularx}

\vspace{2pt}
\noindent\textbf{(b) Run-level parameters for the H$_2$O/cc-pVDZ hardware scan}\par
\vspace{1pt}
\begin{tabularx}{\textwidth}{@{}>{\raggedright\arraybackslash}p{0.235\textwidth} >{\raggedright\arraybackslash}X@{}}
\toprule
Category & Reported value \\
\midrule
{Quantum backend} & IBM Quantum \texttt{ibm\_torino}; 133-qubit Heron r1 processor. \\
Molecular geometry & Symmetric O--H stretch with fixed H--O--H angle $104.485^{\circ}$; $R=0.7,0.8,\ldots,2.0$~\AA\ in increments of 0.1~\AA,
augmented by $R=0.95782$~\AA, for a total of 15 geometries. \\
Sampling & 20,000 shots per submitted ISA (instruction set architecture) circuit and geometry; 300,000 shots over the 15-point scan \\
{State-preparation ansatz} & {Spin-balanced LUCJ sampling circuit initialized from the restricted Hartree--Fock determinant}, with one repetition, $L=n_{\mathrm{reps}}=1$. The LUCJ parameters were obtained from the CCSD amplitudes independently at each molecular geometry; no hardware-side
variational optimization was performed. \\
{Hamiltonian and qubit representation} & {Born--Oppenheimer active-space Hamiltonian and Jordan--Wigner mapping, with one qubit per active spin orbital, $N_q=2N_{\mathrm{orb}}$, and no qubit-count reduction.}\\
Transpilation & Qiskit preset pass manager, optimization level 3, with \texttt{ffsim.qiskit.PRE\_INIT} {inserted in the \texttt{pre\_init} stage}. \\
One-qubit operations & Finite-duration operations $N_{\mathrm{SX}}+N_X=4{,}568$--$5{,}063$; virtual $R_Z$ frame updates $=4{,}018$--$4{,}115$ {(minimum--maximum over the 15 geometry-specific ISA circuits)} \\
Two-qubit operations & $N_{\mathrm{CZ}}=1{,}550$--$1{,}817$ {(minimum--maximum over the 15 geometry-specific ISA circuits)} \\
SQD post-processing & $B=5$ batches {per configuration-recovery iteration}; 350 samples per batch; {$|E^{(k+1)}-E^{(k)}|<2\times10^{-4}\,\sqdEh$ and
 $\max_{p,\sigma}  |n_{p\sigma}^{(k+1)}-n_{p\sigma}^{(k)}|  <2\times10^{-4}$
};
at most 30 recovery iterations and {200 selected-CI solver cycles per batch diagonalization}; random seed 24 \\
{Configuration recovery} & 
Initial spin-resolved particle-number postselection followed by occupancy-guided probabilistic bit-flip 
recovery, repeated postselection, and self-consistent occupation updates. \\
{Final-batch subsampling spread} & Standard deviation across the five final subsamples: 0.0247--0.1029~mHa over the scan and 0.0423~mHa at $R=0.95782$~\AA; {an intra-pool subsampling diagnostic, not an uncertainty estimated from independent quantum-processing unit (QPU) runs} \\
\bottomrule
\end{tabularx}

\vspace{1pt}
\end{table*}
\endgroup
\FloatBarrier

{The CCSD energies and the $t_1$ and $t_2$ amplitudes used to
construct the LUCJ parameters were obtained independently at each
geometry using RHF-based restricted CCSD (RCCSD) as implemented in PySCF, with {the default thresholds of 1 × 10$^{-7} E_h$ for the energy change  and 1 × 10$^{-5}$ for the norm of the amplitude update,   with a maximum of 50 iterations}. All SQD, CASCI, and CCSD values reported in Table~2 include the nuclear-repulsion and frozen-core contributions. PySCF version 2.11.0 was used for the HeH$^+$ and ArH$^+$ calculations,
whereas version 2.12.1 was used for the H$_2$O calculations. The use of different PySCF versions reflects the dates of the respective production calculations; the reported energies were verified to be unchanged to the precision quoted in Table~2.}

\section{Results}\label{sec:3}

Figure~1 presents the ground-state potential energy curves of HeH$^+$, ArH$^+$, and H$_2$O obtained using the SQD method with the 6-31G and cc-pVDZ basis sets, together with reference CCSD results. For HeH$^+$ and ArH$^+$, the comparison provides a direct test of the SQD framework for astrophysically relevant molecular ions, while the inclusion of H$_2$O allows us to assess its performance for a representative polyatomic molecule.

To quantify the agreement between SQD and classical correlated benchmarks more directly, we summarize in Table~2 the ground-state energies obtained near the equilibrium geometries of HeH$^+$, ArH$^+$, and H$_2$O. These values complement the potential-energy curves shown in
Fig.~1 and provide a concise comparison of the equilibrium-region
energies and the same-active-space SQD deviations. 

For HeH$^+$, the SQD results obtained with the cc-pVDZ basis closely reproduce the CCSD potential energy curve across the entire range of internuclear distances considered. In particular, the equilibrium region near $R \approx$ 0.77 {\AA} and the depth of the potential well are accurately captured. The higher SQD energies obtained with the 6-31G basis are consistent
with the reduced flexibility of the smaller basis. However, because
a same-basis CCSD(6-31G) reference is not shown, the basis-set and
SQD-subspace contributions cannot be separated quantitatively. This difference can be understood from the construction of the underlying basis sets. The cc-pVDZ basis belongs to the correlation-consistent family,
which was designed to recover electron-correlation effects
systematically through balanced increases in angular-momentum
functions~~\cite{ccpvdz_basis1,ccpvdz_basis2}.
In contrast, the 6-31G basis is a smaller split-valence Pople-type basis set originally optimized primarily for Hartree-Fock calculations and therefore provides a less flexible description of electron correlation and polarization effects~~\cite{631g_basis}.
As a result, correlated electronic-structure calculations performed with cc-pVDZ generally yield more accurate ground-state energies and potential-energy curves.

A similar trend is observed for ArH$^+$. Despite the larger active space and the heavier atomic species involved, the SQD results with the cc-pVDZ basis again agree closely with the CCSD reference curve throughout the considered bond-length range. The equilibrium region near $R \approx$ 1.27 {\AA} and the overall curvature of the potential energy curve are both reproduced with good accuracy. As in the HeH$^+$ case, the 6-31G results retain the qualitative structure of the potential energy curve but exhibit larger deviations in absolute energy.

The H$_2$O results show that this behavior extends beyond diatomic ionic systems. For the symmetric O--H bond stretching considered here, the SQD potential energy curve obtained with the cc-pVDZ basis follows the CCSD reference closely in the vicinity of the equilibrium geometry and along the scanned bond-length region. Although the 6-31G basis again produces a systematically higher energy, the qualitative shape of the potential energy curve is preserved. This demonstrates that the SQD framework remains effective when applied to a standard polyatomic molecular benchmark, thereby extending its applicability beyond the minimal ionic systems considered in the earlier part of this work.

{At elongated O--H bond lengths, the increasing multireference
character may limit the reliability of the RHF-based CCSD
benchmark. The stretched-region comparison should therefore be
interpreted as consistency with the selected CCSD reference rather
than as an assessment against the exact dissociation curve.}

\FloatBarrier
\begin{figure}[htpb] 
\centering
\includegraphics[width=0.6\linewidth]{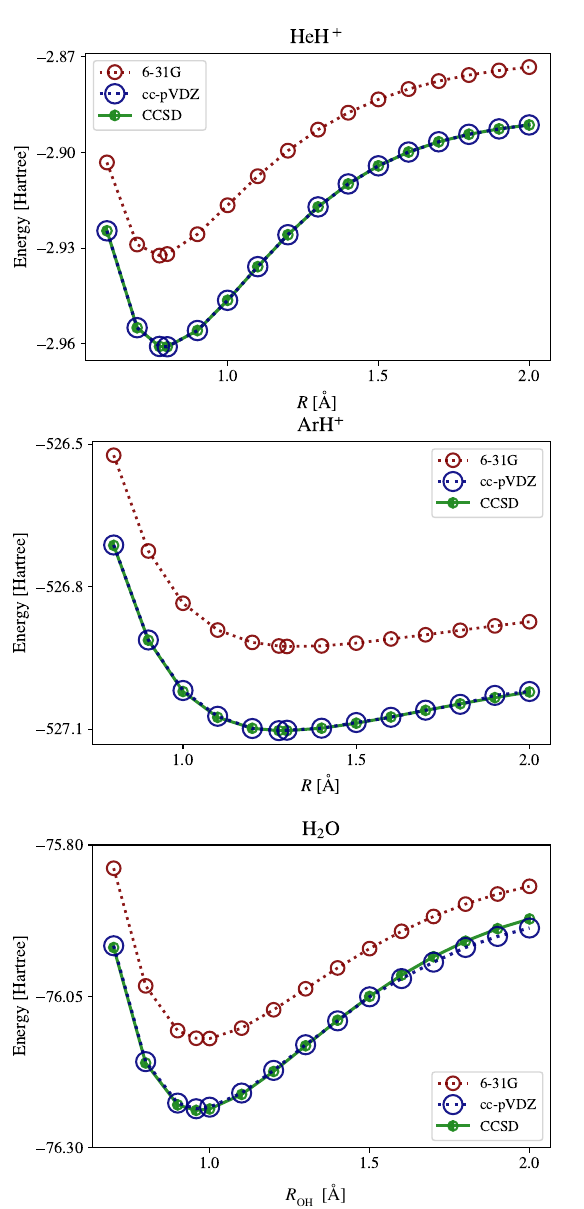}
\caption{{
Ground-state one-dimensional potential-energy curves for
HeH$^+$, ArH$^+$, and H$_2$O obtained using hardware-assisted
SQD with the 6-31G and cc-pVDZ basis sets, together with the
corresponding CCSD/cc-pVDZ references. For the diatomic systems,
$R$ denotes the internuclear distance; for H$_2$O,
$R_{\rm OH}$ denotes the symmetric O--H stretch at the fixed
H--O--H angle of $104.485^\circ$. The SQD/cc-pVDZ curves closely
follow the same-basis CCSD references, particularly near
equilibrium. The SQD/6-31G results preserve the qualitative curve
shapes but give higher total energies. Since CCSD/6-31G curves
are not shown, basis-set incompleteness and SQD-workflow
contributions to these offsets cannot be separated quantitatively.
}}
\label{fig:SQD_HeH_ArH_curves}
\end{figure}

The SQD/cc-pVDZ results closely follow the CCSD/cc-pVDZ reference
near equilibrium. Because a same-basis CCSD/6-31G curve is not
shown, the separate contributions of basis-set incompleteness and
the SQD approximation to the 6-31G offset cannot be quantified.

To quantify the agreement with classical correlated benchmarks more
clearly, Table~\ref{HeH_ArH_SQD_table} reports the equilibrium-region
SQD, CASCI, and CCSD energies obtained with the cc-pVDZ basis.
{Because CASCI diagonalizes the same frozen-core active-space
Hamiltonian used in SQD, $|\Delta E_{\rm CASCI}|$ quantifies the
residual deviation of the complete hardware-assisted SQD workflow
from the exact ground-state energy of that active-space Hamiltonian.
This deviation includes the effects of truncated LUCJ state
preparation, finite-shot hardware sampling, configuration recovery,
and selected-CI subspace construction.} The corresponding quantity
\(
\lvert\Delta E_{\rm CCSD}\rvert
=
\lvert
E_{\rm SQD}^{\rm cc\text{-}pVDZ}
-
E_{\rm CCSD}^{\rm cc\text{-}pVDZ}
\rvert
\)
is reported as a separate comparison with the CCSD benchmark shown
in Fig.~1.
Because the LUCJ parameters were initialized from CCSD amplitudes,
the SQD--CCSD comparison should be regarded as an internal-consistency
benchmark rather than as a fully independent validation.

\FloatBarrier
\begin{table*}[t] 
\centering
\footnotesize
\caption{
Ground-state energies of HeH$^{+}$, ArH$^{+}$, and H$_2$O near their
respective equilibrium geometries, obtained using Sample-Based Quantum
Diagonalization (SQD) with the 6-31G and cc-pVDZ basis sets. The cc-pVDZ CASCI energies are obtained by full configuration-interaction
diagonalization within the same active spaces used in the SQD calculations
and therefore provide the active-space reference used for the convergence
analysis in Fig.~2. The corresponding cc-pVDZ CCSD energies are included as a separate correlated benchmark, consistent with the comparison shown in Fig.~1. The quantities $
\lvert\Delta E_{\rm CASCI}\rvert
=
\lvert E_{\rm SQD}-E_{\rm CASCI}\rvert
$
and $
\lvert\Delta E_{\rm CCSD}\rvert
=
\lvert E_{\rm SQD}-E_{\rm CCSD}\rvert
$
are reported in millihartree (mHa). The literature values are taken from
Refs.~\cite{HeH_ref,ArH_ref,Olsen1996} and are included only as external references because they were obtained using different many-body methods, basis sets, frozen-core conventions, and, in some cases, molecular geometries. {All reported SQD, CASCI, and CCSD energies include the nuclear-repulsion energy and the frozen-core contribution.}
}
\label{HeH_ArH_SQD_table}
\setlength{\tabcolsep}{4.0pt}
\renewcommand{\arraystretch}{1.12}
\resizebox{\textwidth}{!}{%
\begin{tabular}{@{}lcccccccc@{}}
\toprule
\textbf{Molecule}
&
$R$ [\AA]
&
\makecell{\textbf{SQD}\\\textbf{(6-31G)}}
&
\makecell{\textbf{SQD}\\\textbf{(cc-pVDZ)}}
&
\makecell{\textbf{CASCI}\\\textbf{(cc-pVDZ)}}
&
\makecell{\textbf{CCSD}\\\textbf{(cc-pVDZ)}}
&
\makecell{
  $\boldsymbol{\lvert\Delta E_{\mathrm{CASCI}}\rvert}$\\
  \textbf{(mHa)}
}
&
\makecell{
  $\boldsymbol{\lvert\Delta E_{\mathrm{CCSD}}\rvert}$\\
  \textbf{(mHa)}
}
&
\makecell{\textbf{Literature}\\\textbf{$E$ (Ha)}}
\\

\midrule

HeH$^{+}$
&
0.7743
&
{$-2.932301$}
&
{$-2.960790$}
&
{$-2.960790$}
&
{$-2.960790$}
&
{$0.00$}
&
{$0.00$}
&
$-2.9787$~\cite{HeH_ref}
\\

ArH$^{+}$
&
1.2767
&
{$-526.924791$}
&
{$-527.102980$}
&
{$-527.105486$}
&
{$-527.102938$}
&
{$2.51$}
&
{$0.04$}
&
$-527.1827$~\cite{ArH_ref}
\\

H$_2$O
&
0.95782
&
{$-76.118960$}
&
{$-76.235335$}
&
{$-76.241677$}
&
{$-76.238005$}
&
{$6.34$}
&
{$2.67$}
&
$-76.24186$~\cite{Olsen1996}
\\

\bottomrule
\end{tabular}%
}

\end{table*}
The literature energies are included only as external references
because their computational conventions differ from those adopted here.
For the two-electron HeH$^+$ system, SQD, CASCI, and CCSD agree
to the reported precision, giving
$\lvert\Delta E_{\rm CASCI}\rvert=0.00$~mHa and
$\lvert\Delta E_{\rm CCSD}\rvert=0.00$~mHa.
For ArH$^+$ at $R=1.2767$~\AA, the corresponding deviations are
$2.51$ and $0.04$~mHa, respectively.
For H$_2$O at $R_{\rm OH}=0.95782$~\AA, they are
$6.34$ and $2.67$~mHa, respectively.
Thus, the SQD--CASCI deviations quantify the residual deviation of
the complete hardware-assisted SQD workflow from the exact
same-active-space reference, whereas the SQD--CCSD deviations
provide a separate internal-consistency comparison with the CCSD
benchmark.
The literature energies are not used as direct SQD error measures
because their basis sets, molecular geometries, frozen-core
conventions, and correlation treatments may differ from those
adopted here.

At the H$_2$O equilibrium geometry, the final five-batch
subsampling spread is only 0.0423 mHa, more than two orders of
magnitude smaller than the 6.34-mHa SQD--CASCI deviation. The
remaining deviation therefore cannot be attributed to intra-pool
batch fluctuations alone and reflects other components of the
complete SQD workflow, including state preparation, finite-shot
hardware sampling, configuration recovery, and subspace
construction. The batch spread is not, however, an uncertainty
estimated from independent QPU executions.

{The H$_2$O calculation demonstrates that the same hardware-assisted SQD workflow can be applied to a substantially larger, 46-qubit polyatomic active space. The SQD--CASCI deviation, however, increases from 0.00 mHa for HeH$^+$ to 2.51 mHa for ArH$^+$ and 6.34 mHa for H$_2$O. The H$_2$O result should therefore be interpreted as evidence of applicability and feasibility, rather than as evidence of size-independent accuracy or established computational
scalability.} Unlike HeH$^+$ and ArH$^+$, which are both diatomic molecular ions with relatively simple geometrical structure, H$_2$O provides a representative polyatomic test of the SQD framework. Its successful description therefore indicates that the SQD methodology can be applied not only to minimal ionic benchmarks but also to a standard multi-center molecular system. In this sense, H$_2$O serves as an intermediate benchmark between simple diatomic molecules and more chemically realistic larger systems, providing a proof of principle for applying the present SQD workflow to a polyatomic molecular active space. The results indicate that the recovery procedure identifies a
physically meaningful low-energy determinant manifold for both
systems under the present sampling conditions.

\begin{figure}[htbp] 
\centering
\includegraphics[width=0.95\linewidth]{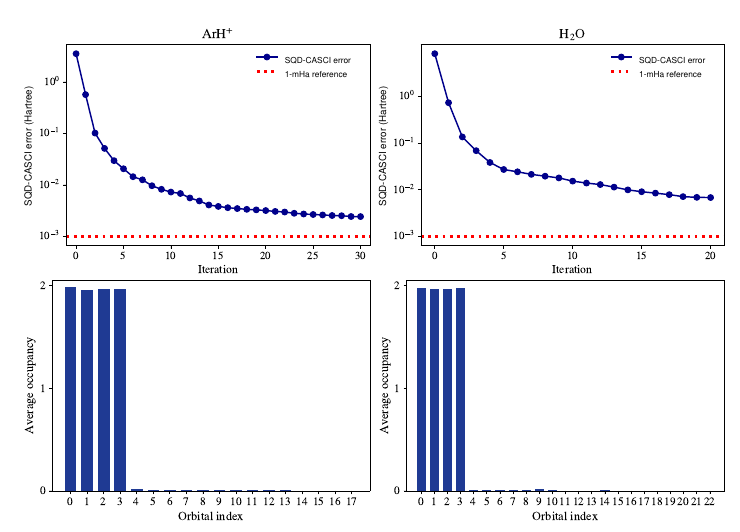}
\caption{
{SQD energy convergence and final orbital occupations} for ArH$^{+}$ (left) at an internuclear distance of
$R=1.2767$~\AA\ and H$_2$O (right) at an O--H bond length of
$R_{\rm OH}=0.95782$~\AA, using the cc-pVDZ basis set.
(Top) Absolute deviation of the SQD energy from the CASCI reference
obtained by full configuration-interaction diagonalization within the
same frozen-core active space,
\(
\delta E_{\rm CASCI}^{(k)}
=
\left|
E_{\rm SQD}^{(k)}
-
E_{\rm CASCI}^{\rm cc-pVDZ}
\right|,
\)
as a function of the configuration-recovery iteration index $k$.
Here,
\(
E_{\rm SQD}^{(k)}
=
\min_{b=1,\ldots,B}
E_{\rm SQD}^{(k,b)}
\)
with $B=5$ batches.
The dashed line denotes the adopted 1-mHa
($10^{-3}$~Ha) reference threshold.
(Bottom) Spin-summed average occupation number,
\(
\bar n_p
=
n_{p\alpha}+n_{p\beta},
\)
for each active spatial orbital at the final recovery iteration,
averaged over the five batch solutions.
The dominant occupations of the lowest spatial orbitals are consistent
with the expected closed-shell ground-state configurations.
}
\label{fig:SQD_ArH_error_avgocc}
\end{figure}
Figure~\ref{fig:SQD_ArH_error_avgocc} illustrates the convergence
of the SQD ground-state energy and the corresponding average orbital
occupations for ArH$^+$ at $R=1.2767$~\AA\ and H$_2$O at
$R_{\rm OH}=0.95782$~\AA.
The upper panels show the absolute deviation from the
same-active-space CASCI reference as a function of the
configuration-recovery iteration index, whereas the lower panels show
the spin-summed average occupation per active spatial orbital at the
final recovery iteration.
The left and right columns correspond to ArH$^+$ and H$_2$O,
respectively.
The final SQD--CASCI deviations are $2.51$~mHa for ArH$^+$ and
$6.34$~mHa for H$_2$O, consistent with
Table~\ref{HeH_ArH_SQD_table}.

{The SQD--CASCI deviation decreases rapidly during the early
configuration-recovery iterations and subsequently approaches a
plateau in the few-millihartree regime. Because all recovery
iterations reuse the same finite 20,000-shot empirical measurement
pool, the deviation is not required to decrease monotonically.
Convergence is therefore determined from the simultaneous
stabilization of the minimum-batch energy and the spin-resolved
orbital occupations, rather than from continued approach to the
CASCI value. The retained H$_2$O run satisfied these criteria after
21 recovery iterations, and no additional post-convergence
iterations were performed.}

The bottom panel provides insight into the final batch-averaged orbital-occupation distribution. In the present calculations, five core orbitals of the Ar atom were treated as frozen and excluded from the active space.
The occupancies therefore correspond to the remaining eight valence electrons distributed among the active spatial orbitals.
The lowest four orbitals exhibit occupancies close to two, consistent with a closed-shell ground-state configuration.

The small but finite occupations of the higher orbitals may reflect
both correlation-induced virtual-orbital population and residual
sampling or recovery effects. Their magnitudes are much smaller than
those of the four predominantly occupied orbitals. The spatial-orbital occupancy was obtained as
$\bar n_p=n_{p\alpha}+n_{p\beta}$ and averaged over the five
final-iteration batches.

Overall, the converged energy trend, together with the physically consistent
final orbital-occupancy distribution, indicates that the SQD framework reliably captures the essential electronic structure of ArH$^{+}$ near equilibrium. A similar convergence pattern is observed for H$_2$O, as shown in the right panels of Fig.~2. 
For H$_2$O, the ground-state energy error decreases steadily with the SQD iteration index and approaches a plateau in the
few-millihartree regime. Although it remains above the adopted 1-mHa reference line over the displayed iteration range, {the observed reduction and subsequent stabilization are consistent with numerical convergence under the stopping criteria adopted in
this work.}

The corresponding average orbital occupancies exhibit a physically consistent closed-shell pattern in the low-lying orbitals, while the small residual population in higher orbitals remains negligible. This behavior is analogous to that found for ArH$^+$ and demonstrates that the SQD framework can identify a compact and physically meaningful low-energy determinant manifold not only for diatomic molecular ions but also for a standard multi-center molecule such as H$_2$O. The broadly similar qualitative convergence trends indicate that the same recovery procedure remains applicable to the larger H$_2$O active space under the present computational settings.

Taken together, the present calculations demonstrate the
feasibility of the hardware-assisted SQD workflow for the selected
molecular benchmarks, including the 46-qubit H$_2$O active space.
For the cc-pVDZ basis, the SQD curves closely follow the same-basis
CCSD references near equilibrium. The increasing SQD--CASCI
deviation indicates, however, that the residual error is not
independent of active-space size. A systematic assessment of
basis-set convergence, robustness, and computational scaling
requires same-basis classical references, independently repeated
hardware executions, and additional systems with larger active
spaces.

\section{Conclusion}

In this work, we applied Sample-Based Quantum Diagonalization (SQD) to the molecular systems HeH$^+$, ArH$^+$, and H$_2$O, and showed that the method can accurately reproduce ground-state potential energy curves and equilibrium-region energetics using shallow-circuit quantum sampling combined with classical subspace diagonalization. With the cc-pVDZ basis, the SQD results agree closely with CCSD benchmarks for all three systems.

The inclusion of H$_2$O is particularly important because it demonstrates that the SQD framework can be extended beyond astrophysically relevant diatomic molecular ions to a representative polyatomic molecular benchmark. This indicates that SQD is not limited to minimal two-center systems, but can also describe a standard multi-center molecule with a larger active space and more complex electronic structure.

Overall, these results demonstrate the feasibility of the present
NISQ-oriented SQD workflow for the selected molecular benchmarks,
including a 46-qubit H$_2$O active space. The calculations provide
a proof of principle for combining hardware-assisted sampling with
classical selected-CI diagonalization in polyatomic systems.
Because the LUCJ parameters were initialized from classical CCSD
amplitudes and each geometry was represented by a single empirical
measurement pool, independently repeated hardware executions,
alternative state-preparation strategies, and systematic
resource-scaling analyses are required before drawing conclusions
regarding robustness, scalability, or end-to-end computational
advantage.

{More broadly, the present results suggest that Sample-Based Quantum Diagonalization may be viewed as a quantum-assisted
configuration-interaction framework in which shallow quantum sampling identifies physically important determinants for
subsequent classical diagonalization.}


	 	

\bmhead{Acknowledgements}
This research was supported by grants from the National Research Foundation of Korea (Grant Nos. NRF-2020R1A2C3006177,
NRF-2021R1A6A1A03043957, and 2018R1D1A1B07051126).
The work of M.-K.C. was supported by the National Research Foundation of Korea (Grant Nos. RS-2021-NR060129 and RS-2025-16071941). The work of J.P. was supported by the ``Quantum Information Science R\&D Ecosystem Creation'' program through the National Research Foundation of Korea (NRF),
funded by the Korean government (Ministry of Science and ICT, MSIT) (Grant No.~2020M3H3A1110365), and by NRF grants funded by the Korean government (MSIT)
(Grant Nos.~RS-2022-NR070836, RS-2025-24533596, and RS-2025-25400847).

\bmhead{Author contributions}
M.-K.C. conceived the study. J.P., C.-H.Y., and M.L. carried out the calculations and data analysis. All authors discussed the results, contributed to the interpretation, and participated in writing and revising the manuscript.

\bmhead{Competing interests}
The authors declare no competing interests.

\bmhead{Data availability}
The data that support the findings of this study are available from the corresponding author upon reasonable request.

\bmhead{Code availability}
The codes used in this work are available from the corresponding author upon reasonable request.

\end{document}